\documentclass[12pt]{article}
\usepackage{graphicx}
\usepackage{dcolumn}
\usepackage{bm}
\begin{document}

\title{Breakdown of supersymmetry in homogeneous cosmologies in N=1 
supergravity}
\author{J\'anos Maj\'ar, and Gyula Bene\\
Institute for Theoretical Physics, 
 E\"otv\"os University\\
     P\'azm\'any P\'eter s\'et\'any 1/A\\ H-1117 Budapest, Hungary\\
E-mail: bene@arpad.elte.hu}
\date{\today}




\maketitle

\abstract{ 
A condition of supersymmetric cosmological solutions
of simple (N=1) supergravity is formulated in the classical case.
As an application we prove that supersymmetry is spontaneously broken
in Friedmann-Robertson-Walker type cosmologies as well as 
in the Kasner universe, except for the Minkowski space. 
} 

\section{Introduction}
Supergravity\cite{supergravity} is actually the gauge theory of supersymmetry\cite{SUSY},
i.e., a theory covariant under local supersymmetry transformations. As the anticommutators of the 
generators of supersymmetry amount to spacetime translations, covariance under local supersymmetry transformations
implies covariance under general coordinate transformations. This means that supergravity 
is a generalization of general relativity, sometimes called the ``square root'' of general relativity.

One motivation behind supergravity theories was the hope that upon quantization a finite perturbation theory
might arise (the zeroth order being the Minkowski spacetime). Clearly, if divergencies arise,
they cannot be removed by renormalization (just like in quantized general relativity), thus a series of miraculous
cancellations is the only possibility to get a meaningful theory\cite{cancellation}. Despite the initial successes,
this expectation proved to be wrong\cite{nocancellation}. 
On the other hand, it turned out that supergravity was a suitable
limiting case of (renormalizable) superstring theories\cite{string} which underlined its possible physical significance.
The situation is in this respect somewhat similar to the relationship between Fermi's theory of weak interactions and
the Weinberg-Salam model. 

If supersymmetry and supergravity indeed plays (or has played) a role in the physical world,
its cosmological consequences might be especially important in the evolution of the early universe.
Indeed, homogeneous supersymmetric cosmological models have been considered in several papers\cite{susy_cosmology}.
The condition of homogenity means invariance of a solution under certain space translations (generated by Killing vectors).
A reasonable generalization can be the condition that a solution is invariant under certain supersymmetry
transformations. If two such supersymmetry transformations exists, their anticommutator defines a suitable
Killing field, hence, homogenity. In the present paper our aim is 
to study the possibility of supersymmetric classical solutions of simple (N=1) supergravity.

\section{The Lagrangian of N=1 supergravity}
We adhere to the conventions of Ref.\cite{supergravity}. The Lagrangian
of simple supergravity is given by
\begin{eqnarray}
\mathcal{L}=-\frac{1}{2}ee^{a\nu}e^{b\mu}R_{\mu\nu
  ab}(\omega)-\frac{1}{2}\epsilon^{\mu\nu\rho\sigma}\bar\Psi_{\mu}\gamma_5\gamma_{\nu}D_{\rho}(\omega)\Psi_{\sigma}
\label{140}
\end{eqnarray}
 
where $e^{a\nu}$ stands for the tetrad field, $\Psi_{\mu}$ for the
gravitino field (having anticommuting spinorial components) and
$\omega$ is a spin connection defined by
\begin{eqnarray}
\omega_{\mu mn}=\omega_{\mu mn}(e)+\frac{\kappa^2}{4}(\bar\Psi_{\mu}\gamma_m\Psi_n
+\bar\Psi_m\gamma_{\mu}\Psi_n+\bar\Psi_m\gamma_n\Psi_{\mu})
\label{127}
\end{eqnarray}
where
\begin{eqnarray}
\omega_{\mu}^{mn}(e)=\frac{1}{2}e_m^{\nu}(\partial_{\mu}e_{n\nu}-\partial_{\nu}e_{n\mu})-\frac{1}{2}e_n^{\nu}(\partial_{\mu}e_{m\nu}-\partial_{\nu}e_{m\mu})-\nonumber\\
-\frac{1}{2}e_m^{\rho}e_n^{\sigma}(\partial_{\rho}e_{c\sigma}-\partial_{\sigma}e_{c\rho})e^c_{\mu}
\label{121}
\end{eqnarray}
The Lagrangian (\ref{140}) changes by a complete divergence 
under the local supersymmetry transformation
\begin{eqnarray}
\delta e^m_{\mu}=\frac{\kappa}{2}\bar{\epsilon}\gamma^m\Psi_{\mu}\label{90}\\
\delta\Psi_{\mu}=\frac{1}{\kappa}\partial_{\mu}\epsilon+\frac{1}{2\kappa}\omega_{\mu}^{mn}\sigma_{mn}\epsilon\label{91}
\end{eqnarray}

\section{SUSY invariance and homogenity}

The supersymmetry transformation is parametrized by the four component 
Majorana spinor field $\epsilon$ whose components are Grassmann variables.
The change of the metric under two successive, local infinitesimal
supersymmetry transformation is given by 
\begin{eqnarray}
\left[\delta_Q(\epsilon_1),\delta_Q(\epsilon_2)\right]g_{\mu\nu}=-\xi_{\mu;\nu}-\xi_{\nu;\mu}-\frac{\kappa^2}{4}\bar\epsilon_2\gamma^{\alpha}\epsilon_1\bar\Psi_{\alpha}\gamma_n\left(\Psi_{\mu}e^n_{\nu}+\Psi_{\nu}e^n_{\mu}\right)
\label{222}
\end{eqnarray}
where
\begin{eqnarray}
\xi^{\mu}=\frac{1}{2}\bar\epsilon_2\gamma^{\mu}\epsilon_1
\label{212}
\end{eqnarray}
Eqs.(\ref{90}), (\ref{91}), (\ref{222}), (\ref{212}) imply that 
if 
\begin{eqnarray}
\delta_Q (\epsilon)e_m^{\mu}=0\label{224}\\
\delta_Q(\epsilon)\Psi_{\mu}=0
\label{225}
\end{eqnarray}
for $\epsilon=\epsilon_1$ and $\epsilon=\epsilon_2$,
then the Lie derivative of the metric along the vector $\xi^{\mu}$ 
vanishes, or, equivalently, $\xi^{\mu}$ satisfies the Killing equation.

\section{Invariance condition for the gravitino field}
Eq.(\ref{225}) can be spelled out as
\begin{eqnarray}
\partial_{\mu}\epsilon+\frac{1}{2}\omega_{\mu}^{mn}\sigma_{mn}\epsilon=0
\label{226}
\end{eqnarray}
or
\begin{eqnarray}
\partial_{\mu}\epsilon=-\frac{1}{2}\omega_{\mu}^{mn}\sigma_{mn}\epsilon
\label{227}
\end{eqnarray}

A necessary condition of solubility of Eq.(\ref{227}) is
\begin{eqnarray}
\partial_{\mu}\partial_{\nu}\epsilon=\partial_{\nu}\partial_{\mu}\epsilon
\label{228}
\end{eqnarray}
Inserting Eq.(\ref{227}) we have
\begin{eqnarray}
\partial_{\mu}\left(-\frac{1}{2}\omega_{\nu}^{mn}\sigma_{mn}\epsilon\right)=\partial_{\nu}\left(-\frac{1}{2}\omega_{\mu}^{mn}\sigma_{mn}\epsilon\right)
\label{229}
\end{eqnarray}
or
\begin{eqnarray}
-\frac{1}{2}\left[(\partial_{\mu}\omega_{\nu}^{mn})\sigma_{mn}\epsilon+\omega_{\nu}^{mn}\sigma_{mn}\partial_{\mu}\epsilon\right]=\nonumber\\
=-\frac{1}{2}\left[(\partial_{\nu}\omega_{\mu}^{mn})\sigma_{mn}\epsilon+\omega_{\mu}^{mn}\sigma_{mn}\partial_{\nu}\epsilon\right]
\label{230}
\end{eqnarray}
By taking $\partial\epsilon$ from (\ref{227}) we get
\begin{eqnarray}
(\partial_{\mu}\omega_{\nu}^{mn})\sigma_{mn}\epsilon+\omega_{\nu}^{mn}\sigma_{mn}\left(-\frac{1}{2}\omega_{\mu}^{ab}\sigma_{ab}\epsilon\right)=\nonumber\\
=(\partial_{\nu}\omega_{\mu}^{mn})\sigma_{mn}\epsilon+\omega_{\mu}^{mn}\sigma_{mn}\left(-\frac{1}{2}\omega_{\mu}^{ab}\sigma_{ab}\epsilon\right)
\label{231}
\end{eqnarray}
or
\begin{eqnarray}
(\partial_{\mu}\omega_{\nu}^{mn}-\partial_{\nu}\omega_{\mu}^{mn})\sigma_{mn}\epsilon+\frac{1}{2}\left[\omega_{\mu}^{mn}\sigma_{mn}\omega_{\nu}^{ab}\sigma_{ab}-\right.\nonumber\\
\left. -\omega_{\nu}^{mn}\sigma_{mn}\omega_{\mu}^{ab}\sigma_{ab}\right]\epsilon=0
\label{232}
\end{eqnarray}
This can be cast to the form
\begin{eqnarray}
\left(\partial_{\mu}\omega_{\nu}^{mn}-\partial_{\nu}\omega_{\mu}^{mn}+\omega_{\mu}^{mc}\omega_{\nu c}^n-\omega_{\nu}^{mc}\omega_{\mu c}^n\right)\sigma_{mn}\epsilon=0
\label{236}
\end{eqnarray}
which is just
\begin{eqnarray}
R_{\mu\nu}^{mn}(\omega(e,\Psi))\sigma_{mn}\epsilon=0
\label{237}
\end{eqnarray}
$R_{\mu\nu}^{mn}$ being the Riemann tensor.
A necessary condition of Eq.(\ref{237}) is that 
\begin{eqnarray}
det\left(R_{(\mu\nu)(mn)}(\omega(e))\right)=0\label{400}
\end{eqnarray}
 
where the components of the Riemann tensor (built exclusively from 
the tetrad field) are ordered into a $6\times 6$ matrix.

\section{Spontaneous breakdown of supersymmetry for homogeneous cosmologies}

Strictly speaking, spontaneous breakdown of a symmetry means that the
ground state of a theory does not possess the symmetry of the Hamiltonian.
Here we mean that a certain kind of solutions (homogeneous
solutions) cannot be invariant under any supersymmetry transformation.
We arrive at this surprising conclusion by applying condition (\ref{400})
for the Riemann tensor calculated from the Friedmann-Robertson-Walker
metric and the metric of the homogeneous and anisotropic Kasner universe.
In the former case we get from (\ref{400}) 
that the second time derivative of the scale factor
vanishes, while in the second case Eq.(\ref{400}) implies that
the metric coincides with the Minkowski metric.
We surmise that this negative result extends to other homogeneous cosmologies
as well. A systematic study of this question is under way.

\section{Conclusion}

We derived the necessary condition (\ref{400}) for a metric to be
the supersymmetric solution of N=1 supergravity. It proved to
be incompatible with the homogeneous Kasner universe and the
Friedmann-Robertson-Walker universe, i.e., a solution of supergravity
of the FRW type cannot be supersymmetric, or in other words,
supersymmetry appears to be spontaneosly broken. This breakdown of
supersymmetry is a consequence of the equations of the theory itself,
especially, no coupling to some hypotetical scalar Higgs 
field has been introduced.

\end{document}